\documentclass[12pt]{iopart}

\usepackage{graphicx}
\begin{document}

\title{Tunnelling in alkanes anchored to gold
electrodes via amine end groups}

\author{Giorgos Fagas and James C. Greer}

\address{
Tyndall National Institute, 	
Lee Maltings,
Prospect Row, 			
Cork, Ireland
}
\ead{
\mailto{georgios.fagas@tyndall.ie},
\mailto{jim.greer@tyndall.ie}
}

\begin{abstract}
For investigation of electron transport on the nanoscale, 
a system possessing a simple to interpret electronic
structure is composed of alkane chains bridging two electrodes via 
end groups; to date the majority of experiments and theoretical 
investigations on such structures have considered thiols
bonding to gold electrodes. Recently experiments show that well
defined molecular conductances may be resolved if the thiol end
groups are replaced by amines. In this theoretical study, we investigate 
the bonding of  amine groups to gold clusters and calculate electron
transport across the resulting tunnel junctions. We find
very good agreement with recent experiments for alkane diamines and 
discuss differences with respect to the alkane dithiol system.  
\end{abstract}

\maketitle

\section{Introduction}

Alkanes bridging metal electrodes form a simple physical realisation of a tunnel
junction, thereby, presenting a model system for both experimental and
theoretical studies regarding the electrical conductivity of molecules.
The height of the energy barrier is given by the
large separation of the highest occupied/lowest unoccupied molecular orbitals
with respect to the Fermi energy, whereas the width of the barrier is
controlled by the number of methylene (-CH$_2$-) groups. In addition, the
contact resistance of the junction may be changed
through the different electron transmission properties of the end groups 
used to bond the alkanes to the metallic electrodes. Such configurations
have been realised experimentally using break junctions and
scanning tunneling microscopy (STM) probes as well as in
conductance measurements through self-assembled
monolayers~\cite{XuT03,SCL03,CZT02Nano,HNZ04,LPH07Nano,Reed,ABL06Nat}.

Experimental and theoretical work to date has focused on the alkane
thiol and alkane dithiol systems bonded between gold electrodes, primarily due to
favourable bonding properties of thiols to gold.
Conductance measurements for these
systems have led to a range of values for the decay constant $\beta$
as determined from the formula
\begin{equation}
G = G_c \exp ( -\beta N),
\end{equation}
with $G$ conductance, $N$ the width of the molecular junction (typically given
either in \AA ngstrom or in number of methylene groups) and $1/G_c$ the junction
contact resistance. Experimental and theoretical values of $\beta$ range 
between 0.5 and 1.0 per methylene group for thiol-anchored
alkanes~\cite{XuT03,SCL03,CZT02Nano,HNZ04,LPH07Nano,Reed,ABL06Nat,KG03NL,JLL04CPL,Evers,FDG06}. The magnitude of the contact resistance
shows a more pronounced spread depending on the experimental technique.
These effects can be related to the uncertainty in the binding site
and molecular conformation as well as to the effect of localised
sulphur states near the Fermi energy~\cite{Evers,TS02PSS,FKE04CPL}.
Recently, Venkataraman {\it et al}~\cite{VKT06} and Chen {\it et al}
~\cite{CLH06} have reported conductance measurements of alkane diamine bonded
between gold electrodes and find a well-defined, narrow distribution in
the molecular conductances, with decay constants and contact resistances
extracted from the two experiments of similar value.
Compared to the thiol-anchored alkane bridges, the properties of
these molecular junctions are much less studied from the theoretical point of
view; in particular for electron transport.

In this study, we first consider bonding of the amine groups
to the metal clusters and discuss possible anchoring through
the amine group -NH$_2$ or its dehydrogenated form -NH.
Then, we present our theoretical predictions for the conductance of the
alkane diamine junctions. Current-voltage characteristics
are calculated by applying a recently developed
transport formalism that relies on the maximum  entropy principle with
application of open system boundary conditions through use of the Wigner
function~\cite{DeG04a}.
For treatment of the junction's electronic structure, a configuration interaction
method~\cite{Gre95,Gre98} is employed thus avoiding many issues surrounding the
use of more approximate electronic structure theory approaches in conjunction
with electron transport~\cite{FDG06,NiR03}. Probing for signatures of differing
end group terminations, we calculate electron transport for terminations
of the form -CH$_2$-NH-Au- and -CH$_2$-NH$_2$-Au-
and compare to recent experimental results~\cite{VKT06,CLH06}.

\section{Computational methods}

\begin{figure}[t]
\includegraphics[width=7cm, height=5cm]{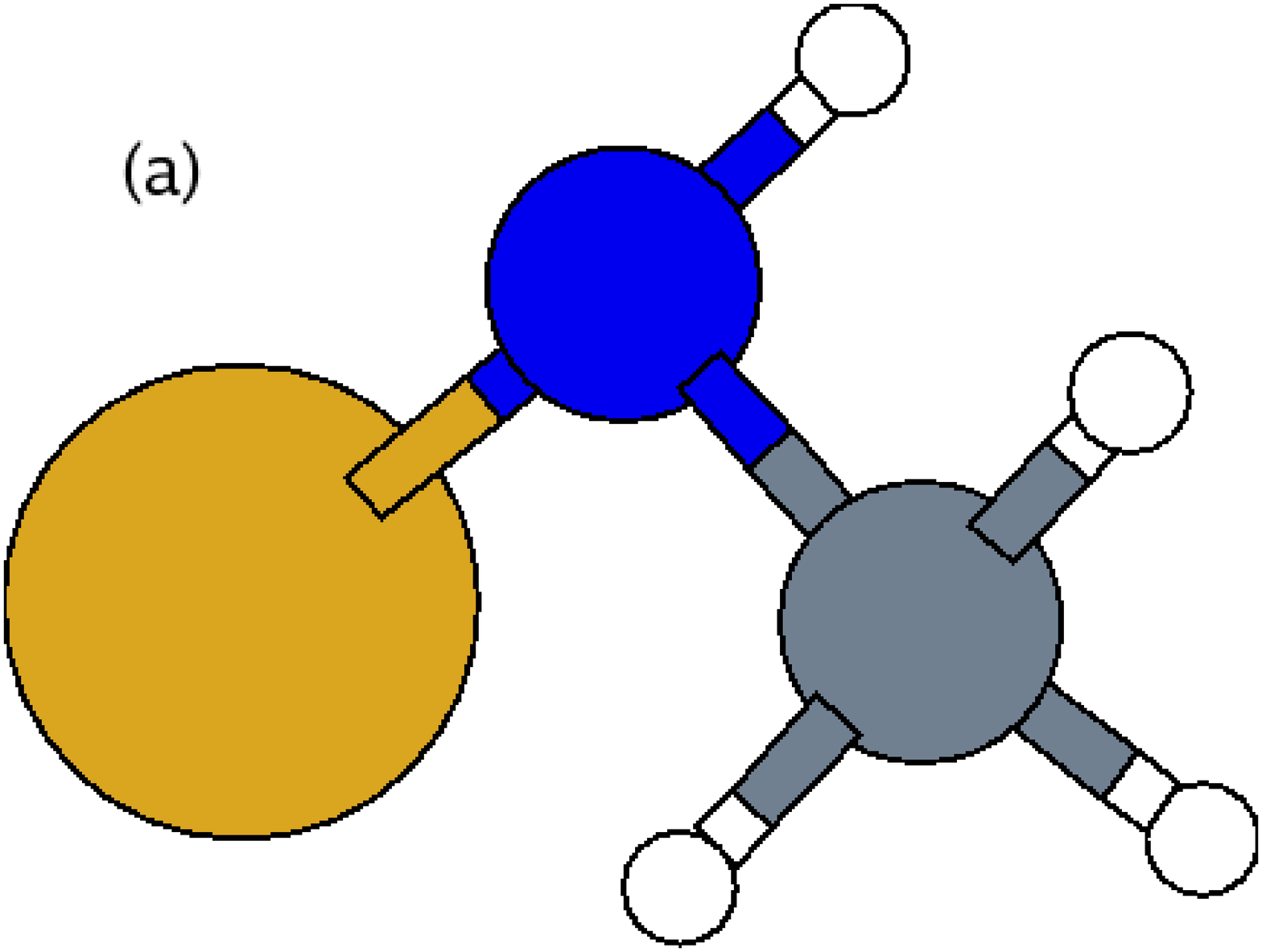}
\\
\vspace{1cm}
\includegraphics[width=7cm, height=5cm]{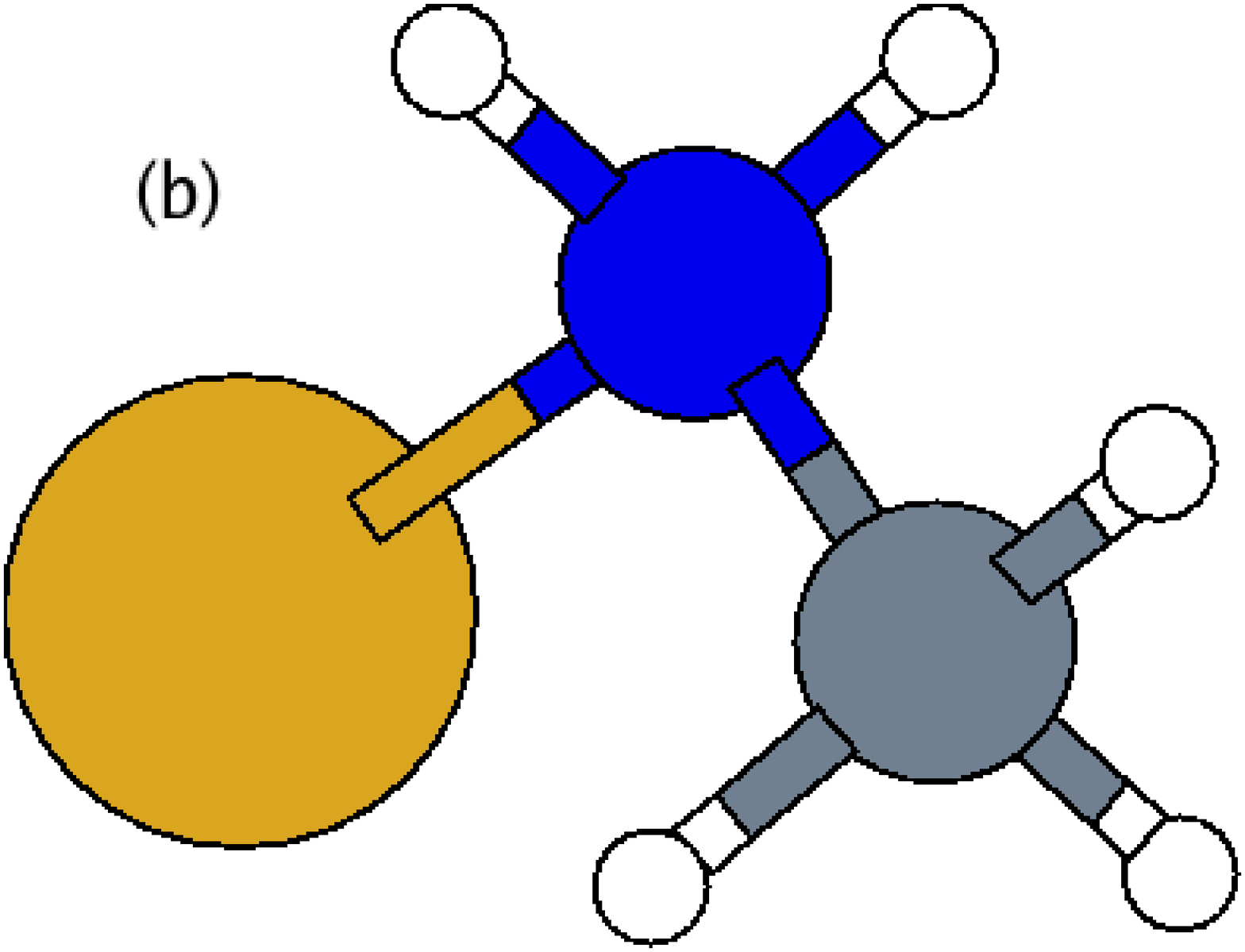}
\caption{Molecular clusters used to investigate bonding of
a) dehydrogenated amine-Au and b) amine-Au.
}
\label{fig1}
\end{figure}

\begin{figure}[t]
\includegraphics[width=9cm, height=7cm]{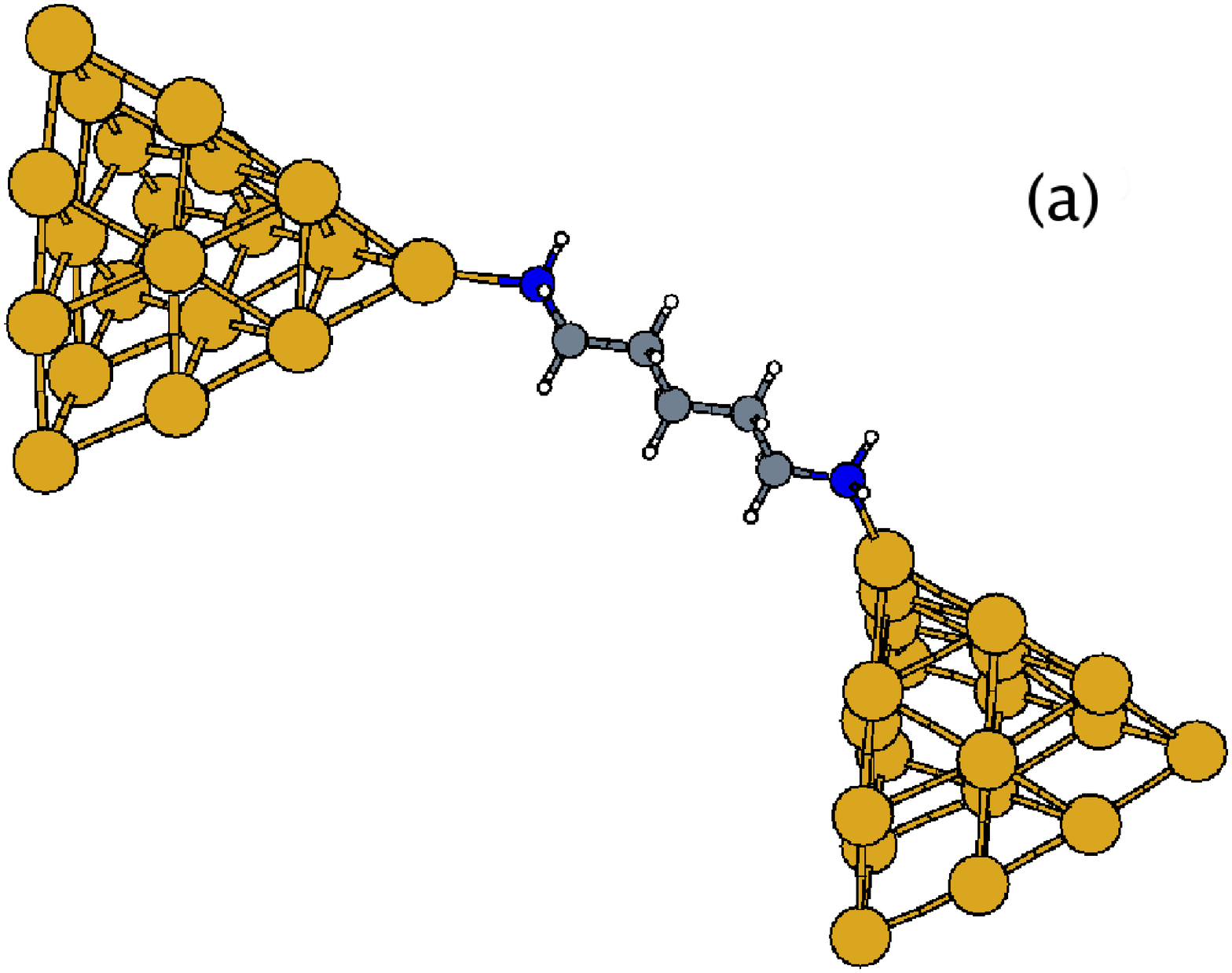}
\\
\vspace{1cm}
\includegraphics[width=9cm, height=7cm]{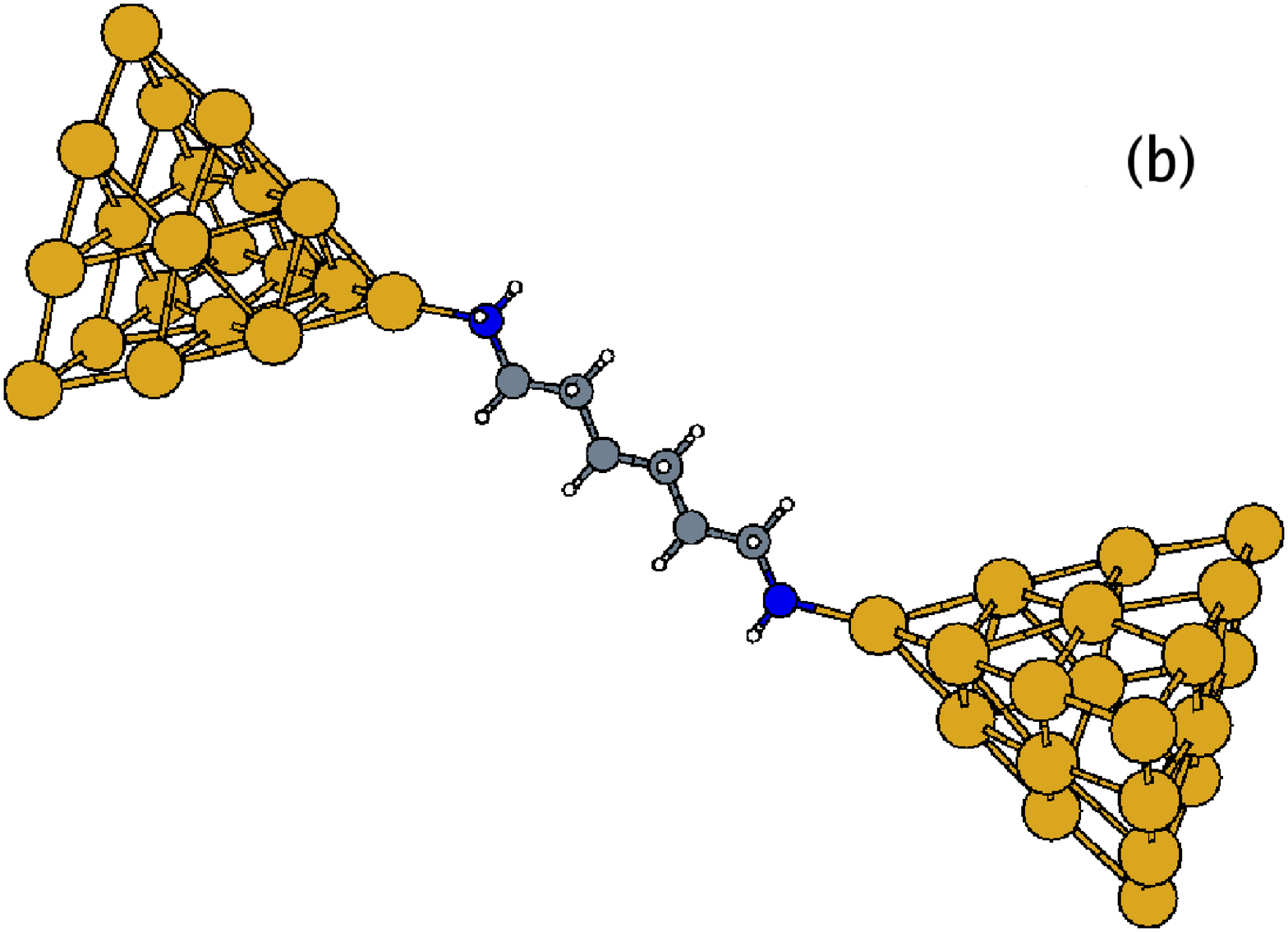}
\caption{Typical metal-molecule-metal optimised structures used in
the current-voltage calculations: a) pentane diamine and b) hexane diamine
bonded to Au$_{20}$ clusters.
}
\label{fig2}
\end{figure}

The energetics of the bonding of amines and their dehydrogenated form
to gold is calculated using density functional theory (DFT)
and Hartree-Fock methods; geometries and energies are reported using DFT
results. For the DFT calculations, we use the B3-LYP hybrid
exchange-correlation functional, for both DFT and Hartree-Fock calculations a
split valence polarised SV(P) basis set is used~\cite{SVP}.
All density functional and Hartree-Fock calculations have been performed
using the TURBOMOLE program package~\cite{ABH89}. For the energy and geometry
calculations, core electrons in gold were removed using an effective core
potential (ECP) leaving 19 electrons per gold atom explicitly
treated~\cite{ECP60}.  Full geometry optimisation
was performed for all molecular clusters considered; see figs.~\ref{fig1}
and \ref{fig2}.

To prepare the many-particle basis set used for the transport 
calculations, the metal electrodes were treated as 20 gold atom clusters
as depicted in fig.~\ref{fig2}. Hartree-Fock calculations were performed on the
DFT optimised structures Au$_{20}$ - NH- (CH$_2$)$_n$ - NH - Au$_{20}$ and 
Au$_{20}$ - NH$_2$- (CH$_2$)$_n$ - NH$_2$ - Au$_{20}$ with $n$=5,6,7,8 or 9
methylene groups. Here, the larger valence double$-\zeta$
correlation-consistent basis set (aug-cc-pvDZ)
was used to treat the carbon atoms~\cite{aug-cc},
whereas, only the gold $6s$ electron was explicitly treated~\cite{ECP-E1}.
Orbitals with an energy less than 15.5 eV above the junction's
highest-occupied molecular orbital were then used in the Monte Carlo
configuration selection procedure~\cite{Gre95,Gre98} with a coefficient
tolerance of 10$^{-3}$. The resulting configuration interaction (CI) vectors
range in length from 5000 to 10000. All transport calculations were performed
with the programme VICI~\cite{DeG04b}.

Our transport formulation is significantly different to common approaches
to electron transport across molecules: for convenience, we outline
the main features of the computations. The quantum chemical
data used to describe the molecular region are subjected to open system boundary
conditions to mimic the action of the electrodes within an experimental
setup. Boundary conditions are imposed using constraints calculated
from the equilibrium density matrix to determine the equilibrium
inward momentum flow from the electrodes, whereas as a voltage is applied the
flow of momentum out of the device region is not constrained. In practice,
the inward and outward momentum flows are defined via the Wigner 
function. Applying these boundary conditions, the
reduced density matrix on the device region (the molecule
plus part of the electrodes) is then calculated at several values of applied 
voltage. The procedure results in the best approximation to the
density matrix on a region subject to reproducing known system observables
in accord with the principle of maximum entropy.
One of the features of our approach is that it allows
for the expansion of many-body states in terms of a complete set of 
many-electron configurations
\begin{eqnarray}
| \Psi \rangle & =&  c_0 | \Psi_0\rangle +
         \sum_{i,a} c_i^a |\Psi_i^a\rangle +
\sum_{i<j}\sum_{a<b} c_{ij}^{ab} |\Psi_{ij}^{ab}\rangle
                  + \ldots
\label{Eq1}
\end{eqnarray}
$| \Psi_0\rangle$ refers to a reference state composed of $N$ lowest single
particle states, $|\Psi_i^a\rangle$ ($|\Psi_{ij}^{ab}\rangle \ldots$) denotes
singly- (doubly- $\ldots$) excited configurations generated by substituting the
$i$-th ($j$-th$, \ldots$) occupied single particle state with the $a$-th
($b$-th$, \ldots$) single particle excitation. 
Notably, our approximation to the device region wavefunction
is not excitation limited and allows for a high 
degree of electron correlation, if required.

\section{Results}

\subsection{Bond analysis}
The bonding of amines to gold is preferentially at adatom
sites~\cite{VKT06}, hence, we first investigate
the bond strength by starting from the simple molecular
clusters AuNH$_2$CH$_3$ and the dehydrogenated form AuNHCH$_3$.
Optimisation of the energy with respect to the atomic positions
results in bond lengths $R_{\rm Au-N}$= 2.03 \AA \, and
$R_{\rm C-N}$=1.46 \AA\, in AuNHCH$_3$ and $R_{\rm Au-N}$=2.38 \AA\,
and $R_{\rm C-N}$=1.46 \AA\, for AuNH$_2$CH$_3$. A strong covalent 
bonding in the dehydrogenated case is reflected in the shorter
gold-nitrogen bond distance $R_{\rm Au-N}$ with
the dehydrogenated amine -NH-. The stronger covalent bonding
for the dehydrogenated linker is also reflected in the bond energies with 
\begin{equation}
\Delta E = E[{\rm AuNHCH}_3]-E[{\rm NHCH}_3]-E[{\rm Au}]=-1.59\, eV,
\end{equation}
whereas the corresponding bond energy for the amine
-NH$_2$ bonding to gold via the nitrogen lone pair is
\begin{equation}
\Delta E = E[{\rm AuNH}_2{\rm CH}_3]-E[{\rm NH}_2{\rm CH}_3]-E[{\rm Au}]=-0.59\, eV.
\end{equation}

However, the relative energies between the two systems
reveal that bonding with the dehydrogenated form is not as stable
as for NH$_2$ with the nitrogen lone pair forming the bond to gold: 
\begin{equation}
\Delta E =
E[{\rm AuNH}_2{\rm CH}_3]-E[{\rm AuNHCH}_3]-\frac{1}{2}E[{\rm H}_2]=-1.03\, eV
\end{equation}

Analysis of the energetics and geometries for the two junctions
Au$_{20}$-NH-(CH$_2$)$_n$-NH-Au$_{20}$ and
Au$_{20}$-NH$_2$-(CH$_2$)$_n$-NH$_2$-Au$_{20}$ yields similar
results as to the simple cluster models. Anchoring via the
dehydrogenated amine gives $\approx 2.07$ \AA \ and $1.45$ \AA \
for the Au-N and N-C bond length, respectively. For
Au$_{20}$-NH$_2$-(CH2)$_n$-NH$_2$-Au$_{20}$, the Au-N and N-C bond length 
are $\approx$ $2.34$ \AA \ and  $1.48$ \AA \, respectively.
In this case too, relative energies of the two systems suggest greater stability
for the Au$_{20}$-NH$_2$-(CH$_2$)$_n$-NH$_2$-Au$_{20}$ junction.
For example, the energy difference between the two anchoring structures
in  hexane- and octane- based molecular junctions is
\begin{eqnarray}
\Delta E &= E({\rm Au}_{20}{\rm NH}_2({\rm CH}_2)_n{\rm NH}_2{\rm Au}_{20}) \nonumber \\
& - E({\rm Au}_{20}{\rm NH}({\rm CH}_2)_n{\rm NHAu}_{20}) - E[{\rm H}_2] \approx -4.9\, eV,
\label{Eq1}
\end{eqnarray}
suggesting that the stronger covalent bond for the dehydrogenated amines causes
distortion to the gold electrode when contact is made.

\subsection{Transport properties}

\begin{figure}[t]
\includegraphics[width=9cm, height=7cm]{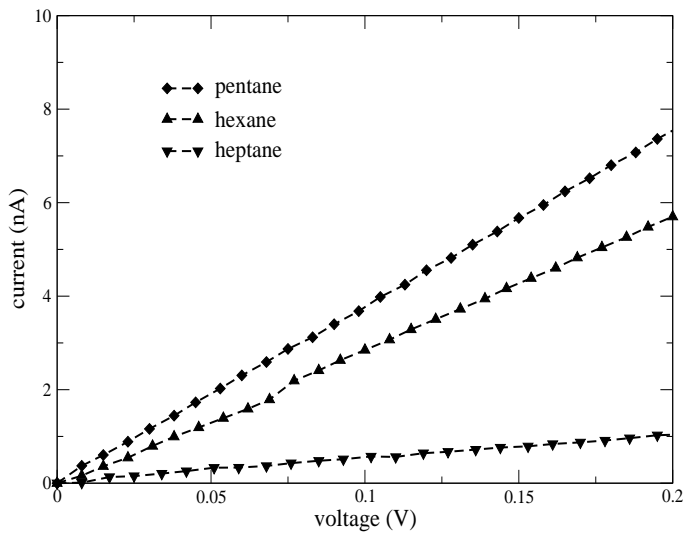}
\caption{Typical current-voltage characteristics for pentane diamine,
hexane diamine and heptane diamine.
}
\label{fig3}
\end{figure}

\Table{\label{tab1}Comparison of measured values
with our predictions for
the conductance of alkane molecules between gold electrodes.
Theoretical results are calculated for NH- and NH$_2$- anchoring groups.}
\br
&\centre{4}{Conductance G(nS)}\\
\ns
&\crule{4}\\
Molecule&NH-anchoring&NH$_2$-anchoring&
L. Venkataraman {\it et al}$^{\rm a}$&F. Chen {\it et al}$^{\rm b}$\\
\ns
&&&&High-G/Low-G\\
\mr
Pentane&$34.52\pm16.51$&$51.99\pm24.10$&$27.12\pm0.77$&-\\
Hexane&$12.96\pm2.98$&$30.64\pm6.07$&$11.62\pm1.16$&$20.79/1.27$\\
Heptane&$4.27\pm1.03$&$4.91\pm3.01$&$5.66\pm1.55$&-\\
Octane&$3.65$&$4.33\pm3.26$&$2.32\pm2.32$&$3.85/0.22$\\
Nonane&-&$1.27\pm0.39$&-&-\\
\br
\end{tabular}
\item[] $^{\rm a}$ Results from the break junction experiments of ref.\cite{VKT06}
\item[] $^{\rm b}$ Results from the STM contact experiments of ref.\cite{CLH06}
\end{indented}
\end{table}

In Table~\ref{tab1}, a comparison of the molecular conductances
obtained from our transport calculations for both amine and dehydrogenated
amine bonding to gold are listed as function of the number of methylene groups
and compared to the recent experimental work of Venkataraman
{\it et al}~\cite{VKT06} and Chen {\it et al}~\cite{CLH06}.
Examples of typical current-voltage characteristics obtained are
shown in fig.~\ref{fig3}.
Overall the agreement with experiment is very good both for the magnitude
of the currents and for decay constants. The measurements yielded
a contact resistance of 430 k$\Omega$ ~\cite{VKT06} and
of 350 k$\Omega$~\cite{CLH06},''high''-G configuration), and,
$\beta$ values of 0.91 $\pm$ 0.03/CH$_2$ and 0.81 $\pm$ 0.01/CH$_2$ were
extracted in ref.~\cite{VKT06} and ref.~\cite{CLH06}, respectively.

For the case of bonding with the -NH$_2$-
we calculate a decay constant $\beta$=0.98 per methylene and a contact
resistance of 140 k$\Omega$. Our previous results for alkane dithiols
yielded a similar $R_C=1/G_C$ but much slower exponential decay of the
conductance with $\beta$=0.5 per methylene~\cite{FDG06}.
On the other hand, bonding through the dehydrogenated linker gives
a theoretical value of $\beta$=0.79 per methylene and a higher contact
resistance of $650$k$\Omega$. Note that higher resistance for stronger bonding
is not uncommon as it is the resulting orbital hybridisation at the contact
that largely determines this value.
Nevertheless, these results do not explain the difference between ``high'' and
``low'' conductance peaks observed in the STM contact experiments~\cite{CLH06}
(see last column of Table~\ref{tab1}).

We should also mention that theoretical
comparisons between the two anchoring groups should be viewed cautiously.
As our CI expansions used for the electronic structure determination are
relatively short, it is likely that the prediction of the HOMO-LUMO gap is not
exact and this will directly influence our predicted decay constant. Also the
finite size of the cluster may not yield the precise energy level alignment.
Additionally, there were numerical sensitivities in the calculated currents
for the alkane diamines that we did not encounter for the alkane dithiols.
These relate to where we choose within the Au$_{20}$ clusters for the application
of the open boundary conditions.
However, for converged calculations the calculated current values at a given
voltage never differed more than plus or minus $75\%$ of the mean value;
actual uncertainties in each case are given in the Table.

Finally, we note that secondary peaks occur in the measurements of Venkataraman
{\it et al}~\cite{VKT06} at slightly higher conductances than for those ascribed
to the amine anchoring. Within the context of our results we cannot exclude the
possibility that primary and secondary peaks can be explained
as different bonding configurations through NH- and NH$_2$- groups
or bonding of -NH$_2$- to different sites.
Energetically the dehydrogenated amine bonding scheme seems unlikely,
but the stronger gold-nitrogen bond for the dehydrogenated linker may be able
to stabilize within the break junction.

\section{Conclusions}

Theoretical calculations for the bonding of alkane diamines to
gold electrodes have been presented exploring also the possibility of anchoring
via dehydrogenated amines.
We find the latter bonding scheme to be energetically 
unfavored, nevertheless, electron transport propertied across
tunnel junctions formed in either case correlate very well with the
experiment. Our results support the amine bonding mechanism
proposed earlier~\cite{VKT06}.
Finally, we note that applying the same theoretical methodology as used to study
alkane dithiols results in very good agreement with the measured
contact resistances and decay constants~\cite{VKT06,CLH06}
when applied independently to the new test case of the alkane diamines.

\section{Acknowledgments}
We are grateful to Fang Chen and Nongjian Tao of Arizona State
University and Latha Venkataraman of Columbia University for providing
us with their measured data.
We thank Simon Elliott for helpful suggestions. Funding from 
Science Foundation Ireland (SFI) is acknowledged.
We would also like to thank the Irish Higher Education Authority (HEA) and
SFI for computational resources provided at Tyndall and through
the Irish Centre for High-End Computing (ICHEC).

\end{document}